\documentclass[prl,superscriptaddress,twocolumn,showpacs]{revtex4}
\usepackage{bm}
\usepackage{graphicx}
\begin{document}

\title{Mass Estimation without using MET in early LHC data}
\author{Z. Kang}
\author{N. Kersting}
\email{nkersting@scu.edu.cn}
\affiliation{Sichuan University, Department of Physics,  Chengdu Sichuan 610065, P.R.C.}
\author{M.J.~White}
\affiliation{The University of Melbourne, School of Physics, Parkville Victoria 3010,
Australia }

\begin{abstract}
Many techniques exist to reconstruct New Physics masses from LHC data,
though these tend to either require high luminosity
$\mathcal{O}(100~{\rm fb}^{-1})$, or an accurate measurement of
missing transverse energy (MET) which may not be available in the
early running of the LHC.  Since in popular models such as SUSY a
fairly sharp, triangular dilepton invariant mass spectrum can emerge
already at low luminosity $\mathcal{O}(1~{\rm fb}^{-1})$, a Decay
Kinematics (DK) technique can be used on events near the dilepton mass
endpoint to estimate squark, slepton, and  neutralino masses  without
relying on MET.  With the first $2~{\rm fb}^{-1}$ of $7$~TeV LHC data
SPS1a  masses can thus be found to 20\% or better accuracy, at least several times better than
what has been taken to be achievable.
\end{abstract}

\pacs{07.05.Kf,12.60.Jv}

\maketitle

We in the particle physics community are naturally eager to glimpse
signs of New Physics in early data from the running of the LHC, most
likely to be first seen in anomalous values of inclusive measurements
of lepton and hadronic jet activity, accompanied by missing energy
from an escaping Dark Matter particle. However, to glean more
quantitative information such as new particle masses will require
fitting data to assumed decay topologies, and a plethora of such mass
reconstruction techniques have accumulated over the
years\cite{Barr}. Typically, these require either large datasets, with
integrated luminosities ${\int \cal L} \sim \mathcal{O}(100~{\rm
fb}^{-1})$, as in endpoint formula
techniques\cite{Bachacou:1999zb,Gjelsten:2004ki}, or precision
measurement of missing transverse energy/momentum (MET), as in the
mT2\cite{Lester:1999tx,Barr:2003rg,Cho:2007qv} and polynomial
methods\cite{Kawagoe:2004rz,Cheng:2007xv,Cheng:2008mg}. However,
accurate MET measurements are not expected to be available in early
LHC data\cite{LHCMET}.

The Decay Kinematics (DK) technique, characterized by the full
reconstruction of events that lie near the endpoint of an invariant
mass distribution for the decay products and where the kinematics are
exactly known, has made its debut recently and proven useful as a mass
estimator in the commonly considered scenario of neutralino pair
production\cite{Kersting:2009ne,Kang}, assuming accurately measured
MET at high luminosity. The purpose of this Letter is to demonstrate
that DK may in fact be the technique of choice in early analysis of certain decay channels, in
particular the well-studied squark-initiated cascade that ends in the
lightest supersymmetric particle (LSP) and Dark Matter candidate, the
lightest neutralino $\tilde\chi_1^0$:
\begin{equation}
\tilde q \to q\tilde\chi_2^0 \to q \tilde\ell^\pm\ell^\mp\to q
\ell^+\ell^-\tilde\chi_1^0.\label{eq:cascade}
\end{equation}
We will show that the DK technique can outperform current methods when one
can rely neither on MET nor large datasets.

---------------------------------------------------------

Consider a collection of N events, each having at least one  pair of isolated
opposite-sign same-flavor (OSSF) leptons plus a high energy jet
($\ell^+ \ell^- j$), which we will assume to have arisen from the
aforementioned squark cascade (\ref{eq:cascade}). For each event, the
following mass-shell constraints hold in the narrow-width
approximation:
\begin{eqnarray}
\label{shelleqs} \label{massq}
(P_n^j + P_n^+ + P_n^-+\!\not\!\!P_n)^2 &=& m_{\tilde q}^2,\\ \label{mass2}
(P_n^+ +P_n^- +\!\not\!\!P_n)^2 &=& m_2^2,\\ \label{masss}
(P_n^- +\!\not\!\!P_n)^2 &=& m_s^2,\\ \label{mass1}
\!\not\!\!P_n^2 &=& m_1^2,\label{eq:lspmass}
\end{eqnarray}
where $P_n^j$, $P_n^\pm$, and $\not\!\!P_n$ are the four-momenta in
the $\mathrm{n^{th}}$ event ($n = 1,\ldots,N$) of the jet, leptons and
the LSP, and where $m_{\tilde q}, m_2, m_1$ and $m_s$ are
abbreviations for the relevant squark, neutralino, and slepton
masses. Note the ambiguity in which lepton is assigned to which
sparticle decay. We will return to this below.

For events that have exactly maximum dilepton invariant mass, {\it
i.e.}\ if
\begin{equation} \label{mll}
(P_n^++P_n^-)^2 = \overline{M}_{\ell\ell}^2 \equiv (m_2^2 - m_s^2) (m_s^2 - m_1^2) /m_s^2,
\end{equation}
then as shown in \cite{Kang} the longitudinal component of the LSP's
three-momentum is subject to a coplanarity constraint:
\begin{equation}
\label{pz}
\not\!P_{nz} =x \not\!\!P_{nx} + y\not\!\!P_{ny},
\end{equation}
where
\begin{equation}
x \equiv \frac{P^+_{nz}P^-_{ny} - P^+_{ny}P^-_{nz} }{P^+_{nx}P^-_{ny} - P^+_{ny}P^-_{nx}}
~{\rm and}~
y \equiv \frac{P^+_{nx}P^-_{nz} - P^+_{nz}P^-_{nx}}{P^+_{nx}P^-_{ny} - P^+_{ny}P^-_{nx}}.
\end{equation}
In the spirit of \cite{Webber},
for each event the constraints (\ref{massq})--(\ref{eq:lspmass}) can be
rearranged into three equations linear in the four components of
$\not\!\!P_n$:
 \begin{eqnarray}
\label{shelleqs} \label{lmassq}
(P_n^j \!+\! P_n^+ \!+\! P_n^-)^2 +2\!\not\!P_n \cdot (P_n^j \!+\! P_n^+ \!+\! P_n^-) \!&=&\!
\Delta^2_{\tilde{q}1},~~\\ \label{lmass2}
(P_n^+ \!+\! P_n^-)^2 + 2\!\not\!P_n \cdot (P_n^+ \!+\! P_n^-) \!&=&\! \Delta^2_{21},~~\\ \label{lmasss}
 2\!\not\!P_n \cdot P_n^- \!&=&\! \Delta^2_{s1}.~~
\end{eqnarray}
with $\Delta^2_{\tilde{q}1} \equiv m_{\tilde q}^2 - m_1^2$,
$\Delta^2_{21} \equiv m_{2}^2 - m_1^2$, and
$\Delta^2_{s1} \equiv m_{s}^2 - m_1^2$.
These, when combined with (\ref{pz}), can be used to solve for
$\not\!P_n$ given a mass hypothesis $(m_{\tilde q}, m_2, m_s, m_1)$.
Thus, for a collection of perfect events arising from (\ref{eq:cascade}), the quantity
\begin{equation} \label{xi}
\xi \equiv \sum_{n=1}^N |\sqrt{\not\!{P}_n^2} - m_1|,
\end{equation}
would be exactly zero if the mass hypothesis were correct. Note
that in this procedure the ambiguity in identifying which lepton comes
from the slepton decay can be removed by explicitly reconstructing the
$\tilde\chi_2^0$ rest frame via the technology of \cite{Kang} and
seeing which lepton is parallel/antiparallel to the LSP. We will use
this below.

Of course, in a real data sample no event will lie exactly at the
kinematic endpoint, and one must settle for a collection of events
within some window of size $\epsilon$ near the dilepton
maximum $\overline{M}_{\ell\ell}$. Experimental effects in the measured momenta will introduce
further smearing on the solution, but it is reasonable to believe that
the correct neighborhood in $(m_{\tilde q}, m_2, m_s, m_1)$-space should still be about that
point which minimizes $\xi$.

The search over this mass space can be greatly simplified by assuming
(\ref{mll}) to hold exactly and using information from estimates of
the upper endpoints of the $q\ell\ell$ and $q\ell$ invariant mass
distributions, $M_{q\ell \ell}$ and $M_{q \ell}$, to indicate the
kinematically allowed regions, as these provide analytical constraints
on the unknown masses\cite{Bachacou:1999zb,Gjelsten:2004ki}.
Even a very crude estimate of these endpoints (call them $\overline{M}_{q\ell \ell}$ and $\overline{M}_{q \ell}$) will
 be sufficient input for the DK technique, as we will see below.

We have tested this method in a realistic Monte Carlo simulation of
the SUSY benchmark point SPS1a\cite{Allanach:2002nj} at low integrated luminosity
($2~{\rm fb}^{-1}$) and center-of-mass energy 7~TeV, as expected from the first first
two years of LHC running. For both signal and background production we
use PYTHIA~6.413\cite{Sjostrand:2006za} interfaced to the fast simulation of a generic
LHC detector, AcerDET-1.0\cite{Richter-Was:2002ch}. For further details of the
simulation, see \cite{Kang}. The cuts used to isolate our signal are:
each event must have at least two hard jets with $p_T>150,\,100$~GeV
(expected from pair production of squarks or gluinos), plus two
isolated OSSF leptons ($e^+e^-$ or $\mu^+\mu^-$) with $p_T>10$~GeV. We
apply $p_T$ dependent lepton efficiencies as in\cite{Kang}, based on
full simulation results published in\cite{:2008zzm}.
The jet resolution at $7$~TeV is not yet well understood,
and one possible parametrization of this is to introduce additional smearing of
jet momenta by hand; yet the uncertainty in jet direction is likely to be much smaller
than of jet energy, so it is perhaps more realistic to
scale all components of jet 4-momenta by the same factor ---
 we thus scale all jet energies up by a generous 5\% (scaling down gives virtually the same influence on final results).
Standard Model backgrounds for the jet+dilepton endstate considered are overwhelmingly dominated by $t\overline{t}$  which we have
generated
for this particular machine energy and luminosity (PYTHIA $\sigma_{LO} = 85$~pb), giving $1.7\cdot 10^5$ background events; lesser backgrounds include $ZZ^*$ and $Z+jets$
which are several times smaller and can be essentially eliminated by selecting events away from the Z-pole.
All of this is against
only $1.2\cdot 10^4$ SUSY events (all processes), but we will nevertheless see below that DK has excellent background
rejection.

\begin{figure}
	\begin{center}
	\includegraphics[width=0.45\textwidth]{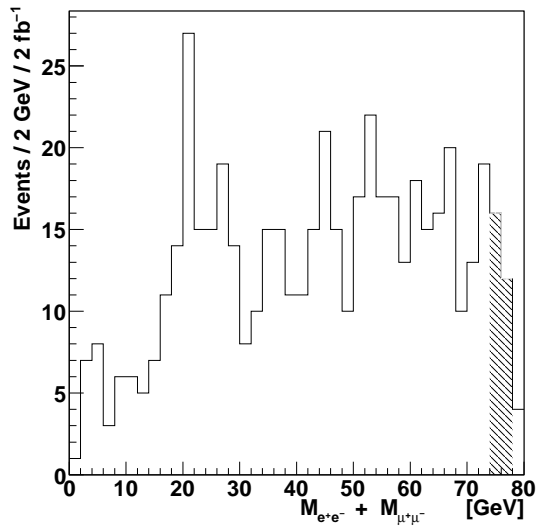}
	\end{center}
	\caption{\label{fig:dilep} Events from the shaded bins of the OSSF distribution ($74~\mathrm{GeV} < M_{\ell \ell} < 78~\mathrm{GeV}$)
are used for
DK analysis.}
\end{figure}

As can be verified from Fig.\ref{fig:dilep}, even at this low
luminosity and set of cuts the OSSF dilepton distribution typically contains hundreds of
events (here about 600, half of which are from SUSY), and dozens of events within a moderately small window (here
chosen to be $\epsilon = 4$~GeV) near the kinematic endpoint at
$\overline{M}_{ll} \approx 77$~GeV.
Note that although backgrounds outnumber signal in general, this is not so for events near the dilepton endpoint:
in this sampling area shown with $74~\mathrm{GeV} < M_{\ell \ell} < 78~\mathrm{GeV}$, 24 of the events are from SUSY while only 4 are
background (S/B = 6), hence choosing events near this endpoint gives us quite some
leverage against backgrounds. In fact, owing to the fact that background events tend to fail out of the DK algorithm (see below), we find that backgrounds only become a serious issue when $S/B \sim 1$ and $S<10$.
We must mention that one has some freedom in choosing $\epsilon$,
balancing the tension between accurate results with small systematic
errors (small $\epsilon$) and low statistical error (larger
$\epsilon$). As a rough criterion in general, one should choose $\epsilon$ so as to give $\mathcal{O}(10)$-$\mathcal{O}(100)$ events;
our experience is that this suffices to give a decent mass estimate.

\begin{figure}
	\begin{center}
	\includegraphics[width=0.40\textwidth]{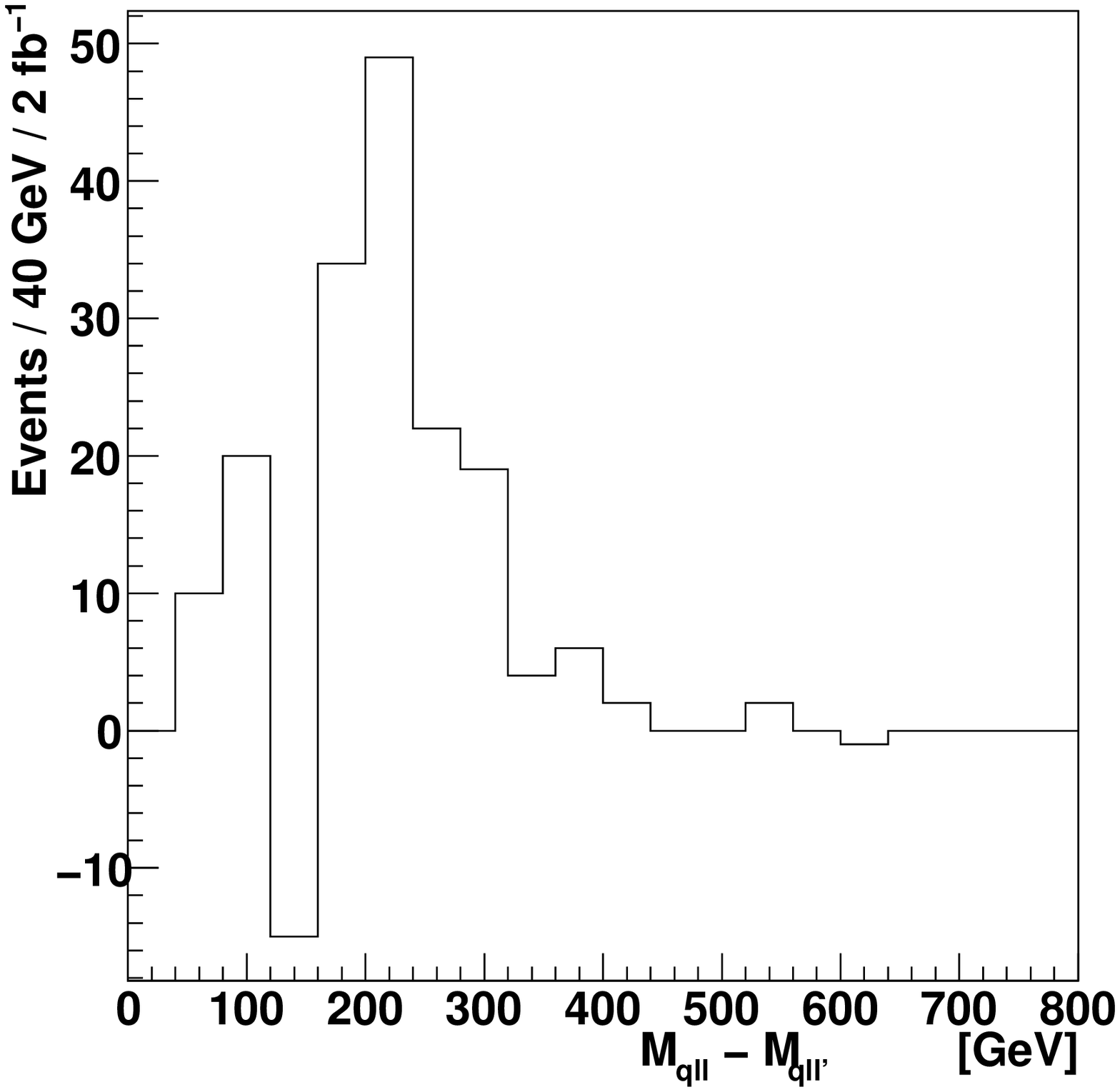}
  \includegraphics[width=0.40\textwidth]{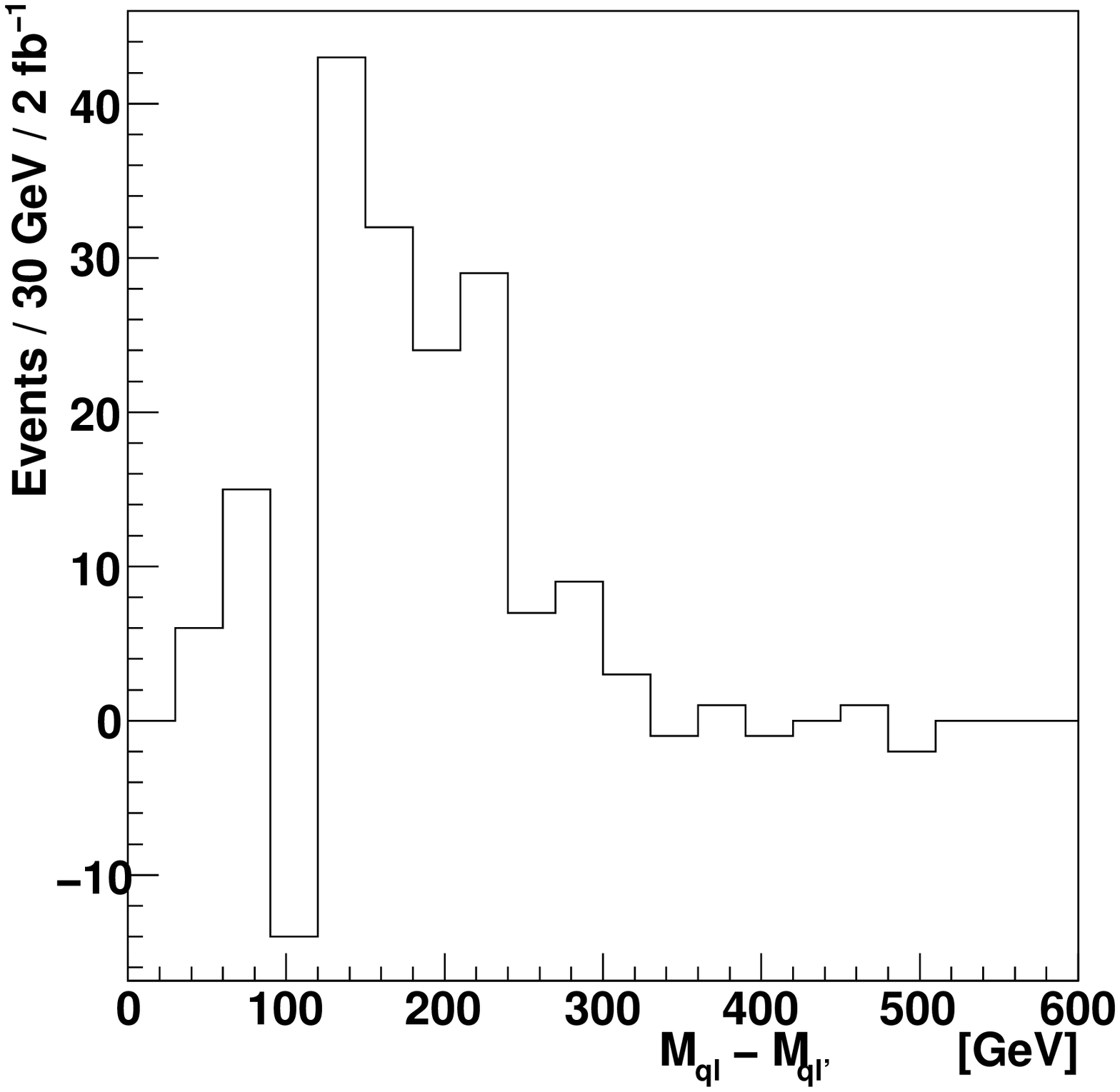}
	\end{center}
	\caption{\label{fig:dists} $M_{q\ell}$ and $M_{q\ell\ell}$ mass distributions, flavor-subtracted as described in the text and in
\cite{Bachacou:1999zb}. A rough estimate of the endpoints suffices for our analysis:  $330~\mathrm{GeV} < \overline{M}_{q\ell\ell} < 430~\mathrm{GeV}$,  $300~\mathrm{GeV} < \overline{M}_{q\ell} <
400~\mathrm{GeV}$.}
\end{figure}

To determine what set of SUSY masses is most consistent with this
sample of events, we perform the following scan over mass space: $m_s$
is chosen in a liberal range, $100~\mathrm{GeV} < m_s < 500~\mathrm{GeV}$ (lower bound from LEP constraints) as well as
$\overline{M}_{q\ell \ell}$ and $\overline{M}_{q \ell}$ in ranges estimated from Fig.\ref{fig:dists},
$300~\mathrm{GeV} < \overline{M}_{q\ell} < 400~\mathrm{GeV}$, $330~\mathrm{GeV} < \overline{M}_{q\ell\ell} < 430~\mathrm{GeV}$.
Note that,
as is customary in handling jet ambiguity, what we have actually plotted are
the minimum values of $M_{q\ell \ell}$ and $M_{q \ell}$ computed from each of the two highest energy
jets; also, before this minimization $M_{q \ell}$ is taken to be the larger of the two possible $q \ell$ masses.
To reduce background and combinatorial effects further, these distributions are also computed for OSOF lepton events,
labeled $M_{q\ell \ell'}$ and $M_{q \ell'}$, and are subtracted from the
$M_{q\ell \ell}$ and $M_{q \ell}$ distributions, respectively.
The total procedure thus improves endpoint precision at the cost of distorting distribution shape
 somewhat, but this is irrelevant to our present purpose.

For each choice of the endpoints $\overline{M}_{\ell \ell}$, $\overline{M}_{q\ell \ell}$ and $\overline{M}_{q \ell}$,  the
other unknown masses are then fixed from analytic expressions of the
endpoints in terms of these masses and $m_s$ as
\begin{eqnarray} \label{m1eqn}
m_1 &=& \sqrt{m_s^2 - m_s \overline{M}_{\ell\ell} \sqrt{\overline{M}_{q\ell\ell}^2/\overline{M}_{q\ell}^2 - 1}},\\ \label{m2eqn}
m_2 &=& m_s\sqrt{1 + \overline{M}_{\ell\ell}^2/(m_s^2 - m_1^2)},\\ \label{mqeqn}
m_{\tilde q} &=& m_2\sqrt{1 + \overline{M}_{q\ell\ell}^2/(m_2^2 - m_1^2)},
\end{eqnarray}
and for each such point in mass space the linear system of equations
(\ref{pz})--(\ref{lmasss}) is solved event-wise for $\not\!P_n$,
constructing the sum (\ref{xi}) defining $\xi$. Both possible lepton
assignments are tried. In this procedure each event contributes the
minimum value of $|\sqrt{\not\!{P}_n^2} - m_1|$ resulting from
matching the OSSF lepton pair to each of the two highest energy jets,
 accepting only solutions where the assumed
lepton identity does indeed lead to reconstruction of the correct
leptonic directions relative to the LSP momentum, i.e. the cosine of the angle between the  assumed near(far) lepton momentum
and the LSP must be  $>\!\!(<) 0$  in the rest frame of the decaying $\tilde\chi_2^0$. If, for any event
$n$, the quantity $|\sqrt{\not\!{P}_n^2} - m_1|$ exceeds a tolerance $T$ (we take $T =30$~GeV), or
if $\not\!{P}_n$ is tachyonic, or if the reconstruction of the
$\tilde\chi_2^0$ velocity fails for both possible jet pairings, the
event is assigned a fixed contribution of $T$ to $\xi$ ({\it
i.e.}\ a fit which is $T = 30$~GeV in error is deemed as bad as finding no fit at
all). As with the $\epsilon$ parameter, $T$ is adjustable and could be  optimized somewhat, but
  any reasonable value will do. The
choice of $(m_s, \overline{M}_{q\ell\ell}, \overline{M}_{q\ell})$ giving the minimum $\xi$
is then assumed to yield the correct mass hypothesis. For the particular data set shown, we obtain
$(m_s, \overline{M}_{q\ell\ell}, \overline{M}_{q\ell}) = (133, 408, 300)$~GeV as the $\xi$-minimizing solution, and using
(\ref{m1eqn})-(\ref{mqeqn}) yields $(m_1, m_s, m_2, m_{\tilde q}) = (90.3,133.0,170.0,510.7)$~GeV,  quite
close to the nominal values of  $(96,143,177,537)$~GeV.

\begin{figure}
	\begin{center}
	\includegraphics[width=0.40\textwidth]{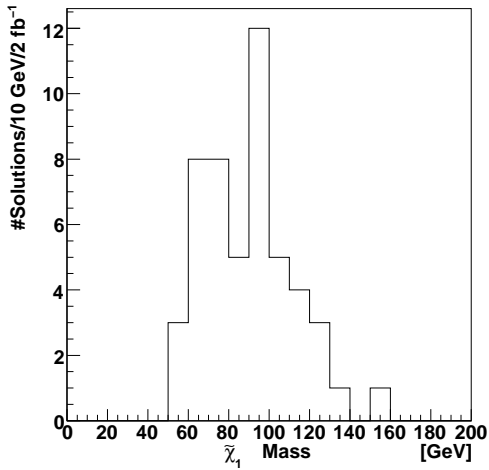}
	\end{center}
	\caption{\label{fig:50trials} Best fits to the LSP mass for each of 50 independently-generated $2~\mathrm{fb}^{-1}$ data
sets (SUSY signal + SM background).}
\end{figure}

To get some idea on the expected accuracy of this method owing to
random fluctuations in signal and background, we look at the best-fit solutions
in 50 independently-generated data sets, obtaining the following
results:
\begin{eqnarray} \label{3fbmasses}
m_1 &=& ~89.9  \pm 21.8~\mathrm{GeV} \\ \nonumber
m_s &=& 131.0 \pm 22.2~\mathrm{GeV} \\ \nonumber
m_2 &=&  169.2 \pm 21.6~\mathrm{GeV} \\ \nonumber
m_{\tilde q} &=& 512.4 \pm 31.4~\mathrm{GeV}
\end{eqnarray}
Quoted uncertainties are actually square-roots of sample variances.
Though these results are slightly lower than the nominal values,  one generally gets accurate
 (within $\sim 15-20\!$~\%) estimates of
the masses, and the spread in the solutions is consistent with the rough width of the peak in, for example, the
LSP mass distribution in Fig.~\ref{fig:50trials}. This width scales with the jet energy-bias we put in by hand, so
once low-luminosity jet reconstruction efficiency is better understood the accuracy of the method would be closer
to $\sim 10-15$\% on average.

Let us again stress that, prior to this Letter,  the only known method of reconstructing particle masses in early-LHC data
was the standard
endpoint analysis, and this will give considerably worse results: in addition to the three endpoints  $\overline{M}_{\ell\ell}$,
$\overline{M}_{q\ell\ell}$, and $\overline{M}_{q\ell}$  whose precise determination is not possible with such low statistics, one needs a
fourth endpoint, usually taken as the lower endpoint of the $M_{q \ell \ell}$ distribution, to solve for a point in $(m_{\tilde{q}}, m_2,
m_s, m_1)$-space. However, this lower endpoint, which is supposed to be at ${M}_{q\ell\ell}=202$~GeV, is not discernible in
Fig.~\ref{fig:dists}. Studies with more sophisticated detector simulation and  endpoint fitting algorithms point to uncertainties in the
masses that can reach $50$\% or more from early LHC data\cite{LHCMET} at comparable luminosities.

---------------------------------------------------------

This Letter demonstrates that, in the early stages of the LHC, one may
obtain far more accurate estimates of New Physics masses than
previously thought if one uses a DK technique, at least for $\ell^+
\ell^-j$ events assumed to arise from the squark cascade chain
considered here.
Generalization to the same final state arising from
other models (such as extra dimensions or little higgs models), or to
different decay topologies is straightforward. What makes DK so
powerful in the present case is not only that one can utilize the high
statistics in the dilepton endpoint, but that one can resolve lepton ambiguity
 through explicit reconstruction that does not rely on
a measurement of MET. In fact, the same method may be used
on a decay chain where the LSP decays through R-parity violating
operators, as there is no reliance on pair-production of new sparticles. The need for an effective mass estimator in the early
stages of the LHC is attractive not only to help theorists hone in on
likely models for New Physics signals, but also in the design and
optimization of experiments, especially with respects to the LSP,
whose mass measurement at the LHC will directly affect designs of Dark
Matter detection facilities elsewhere.

\bigskip

\begin{acknowledgments}
\emph{Thanks to S.~Kraml for some useful suggestions, and to A.~Raklev for helping with background data and giving much invaluable
critique of the manuscript. MJW is supported by the Australian Research Council.}
\end{acknowledgments}

\end{document}